\documentclass[10pt]{iopart}

\newcommand{\onlinecite}[1]{\hspace{-1 ex} \nocite{#1}\citenum{#1}}
\usepackage{iopams}
\usepackage[sort&compress,numbers]{natbib}

\usepackage{graphicx}
\graphicspath{{./figs/}}
\usepackage[utf8x]{inputenc}
\begin{document}

\title[Flux-driven Josephson parametric amplifier for sub-GHz frequencies]{Flux-driven Josephson parametric amplifier for sub-GHz frequencies fabricated with side-wall passivated spacer junction technology}

\author{Slawomir Simbierowicz$^1$, Visa Vesterinen$^{1,2}$, Leif Gr\"onberg$^1$, Janne Lehtinen$^1$, Mika Prunnila$^1$ and Juha Hassel$^1$}
\address{$^1$ VTT Technical Research Centre of Finland Ltd \& QTF Centre of Excellence, P.O.Box 1000, FI-02044 VTT, FINLAND}
\address{$^2$ QCD Labs, COMP Centre of Excellence, Department of Applied Physics, Aalto University, P.O. Box 13500, 00076 Aalto, Finland}
\ead{slawomir.simbierowicz@vtt.fi}
\vspace{10pt}
\begin{indented}
\item[]May 2018
\end{indented}

\begin{abstract}
We present experimental results on a Josephson parametric amplifier tailored for readout of ultra-sensitive thermal microwave detectors. In particular, we discuss the impact of fabrication details on the performance. We show that the small volume of deposited dielectric materials enabled by the side-wall passivated spacer niobium junction technology leads to robust operation across a wide range of operating temperatures up to 1.5~K. The flux-pumped amplifier has gain in excess of 20~dB in three-wave mixing and its center frequency is tunable between 540~MHz and 640~MHz. At 600~MHz, the amplifier adds 105~mK~$\pm$~9~mK of noise, as determined with the hot/cold source method. Phase-sensitive amplification is demonstrated with the device.
\end{abstract}


\maketitle

\ioptwocol

\section{Introduction}

In recent years, high-fidelity detection of radio-frequency (rf) and microwave signals that can consist of only a few photons has spun a lot of interest in the development of low-noise amplifiers. Such weak signals are encountered for instance in the search for dark-matter particles \cite{bradley_microwave_2003, asztalos_squid-based_2010, kenany_design_2017}, fast readout of quantum bits (qubits) \cite{bergeal_phase-preserving_2010, abdo_josephson_2014, obrien_towards_2016, devoret_superconducting_2013}, and characterization of low-loss resonators \cite{calusine_analysis_2018} or nano-mechanical systems \cite{clark_observation_2016}. 
A promising branch of superconducting amplifiers, with near quantum-limited noise performance, exploits parametric pumping of the non-linear inductances exhibited by Josephson junctions \cite{yurke_observation_1988, castellanos-beltran_widely_2007, yamamoto_flux-driven_2008, mutus_design_2013, eichler_controlling_2014, mutus_strong_2014, eichler_quantum-limited_2014} or those intrinsic to superconductors \cite{vissers_low-noise_2016}. The Josephson parametric amplifier (JPA) has also proven to be capable of generating and using squeezed electromagnetic states \cite{mallet_quantum_2011, fedorov_displacement_2016, bienfait_magnetic_2017} to go below the standard quantum limit of noise added by an amplifier \cite{clerk_introduction_2010}.

Although the most common applications for the JPA are in the frequency band of 4--8~GHz, we recently reported on a JPA for 600~MHz~\cite{vesterinen_lumped-element_2017} to be used in conjunction with a nano-calorimeter \cite{gasparinetti_fast_2015, viisanen_incomplete_2015} or a \mbox{-bolometer} \cite{govenius_microwave_2014, govenius_detection_2016} with a matching readout frequency. Our main motivation to develop the JPA is to allow the calorimeter to reach the accuracy of a single microwave photon and set a new record for the noise-equivalent power going below $10^{-19}$~$\mathrm{W}/\sqrt{\mathrm{Hz}}$ \cite{karasik_demonstration_2011, suzuki_performance_2014} in the bolometric mode. More recently, rf reflectometry of charge qubits has also emerged as a possible use case for the sub-GHz JPA \cite{penfold-fitch_microwave_2017}. Aiming to serve such applications, the realized JPA 
utilized the non-linearity of niobium-based superconducting quantum interference devices (SQUIDs) in a lumped-element rf resonator. The amplifier of Ref.~\onlinecite{vesterinen_lumped-element_2017} was narrowband, but the center frequency of the gain was designed to be tunable with an external magnetic flux. However, the device suffered from multiple issues that prohibited its immediate use in calorimetry.

The first prominent issue discovered in Ref. \onlinecite{vesterinen_lumped-element_2017}   was an ill-behaved, hysteretic response of the resonance frequency to the applied magnetic flux. Its origin was attributed to flux trapping in the device geometry. The second issue was a high sensitivity to changes in the operating temperature, requiring stabilization of the JPA with closed-loop temperature control. We believe that the temperature sensitivity stemmed from two-level systems (TLSs) in a deposited dielectric layer that had a large participation ratio to the JPA resonance. This layer was made of silicon dioxide that is notorious for its high TLS density which significantly affects material properties at millikelvin temperatures \cite{oconnell_microwave_2008, pappas_two_2011}.

Here, we seek to improve on the shortcomings of Ref.~\onlinecite{vesterinen_lumped-element_2017} while keeping the design conceptually similar. We present several important modifications to the JPA, the first of which is the fabrication of the Josephson junctions with the so-called side-wall passivated spacer (SWAPS) process that we introduced recently \cite{gronberg_side-wall_2017}. This enables us to largely avoid using plasma-enhanced chemical vapour deposited (PECVD) silicon dioxide which is a necessity in our standard niobium tunnel junction processes \cite{kiviranta_multilayer_2016}. Measures to control the flux trapping are implemented as well. The new JPA also employs three-wave mixing with an rf flux pump at twice the signal frequency \cite{lahteenmaki_dynamical_2013} as opposed to the four-wave mixing of Ref.~\onlinecite{vesterinen_lumped-element_2017} which utilized an rf current pump \cite{castellanos-beltran_widely_2007,mutus_design_2013} in the vicinity of the signal frequency. We report on good measured performance in both the non-degenerate and degenerate modes \cite{devoret_introduction_2016} of the JPA, warranting its later integration into the nano-calorimetry setup.

\section{Devices}

Amplifying the readout signals of a nano-calorimeter requires sufficiently high dynamic range and enough bandwidth from the JPA. More specifically, the amplifier should be able to handle input signals at a maximum power of \mbox{-120}~dBm without going into saturation, and it needs to respond to detector signals in the time scales of 10--1000~$\mu$s. These targets were met in Ref.~\onlinecite{vesterinen_lumped-element_2017} using a JPA realized with a lumped-element LC resonator for radio frequencies. The inductance originated largely from Josephson junctions in a series array of 200 SQUIDs with a maximal critical current of 35~$\mu$A. Shunted with a capacitance of $\simeq$~30~pF, the flux-tunable resonator had a maximum resonance frequency $f_0$ of 650~MHz. The capacitive coupling to an external 50-$\Omega$ rf environment set the gain-bandwidth product to \mbox{$2\pi f_0/Q_e \simeq 2\pi \times 2.2$~MHz}, where \mbox{$Q_e \simeq 300$} is the external quality factor. In this work, the device parameters are similar, and a detailed listing is provided in the Supplement \cite{noauthor_supplementary_nodate}.

\begin{figure*}[htpb]
\centering
\includegraphics[width=\hsize]{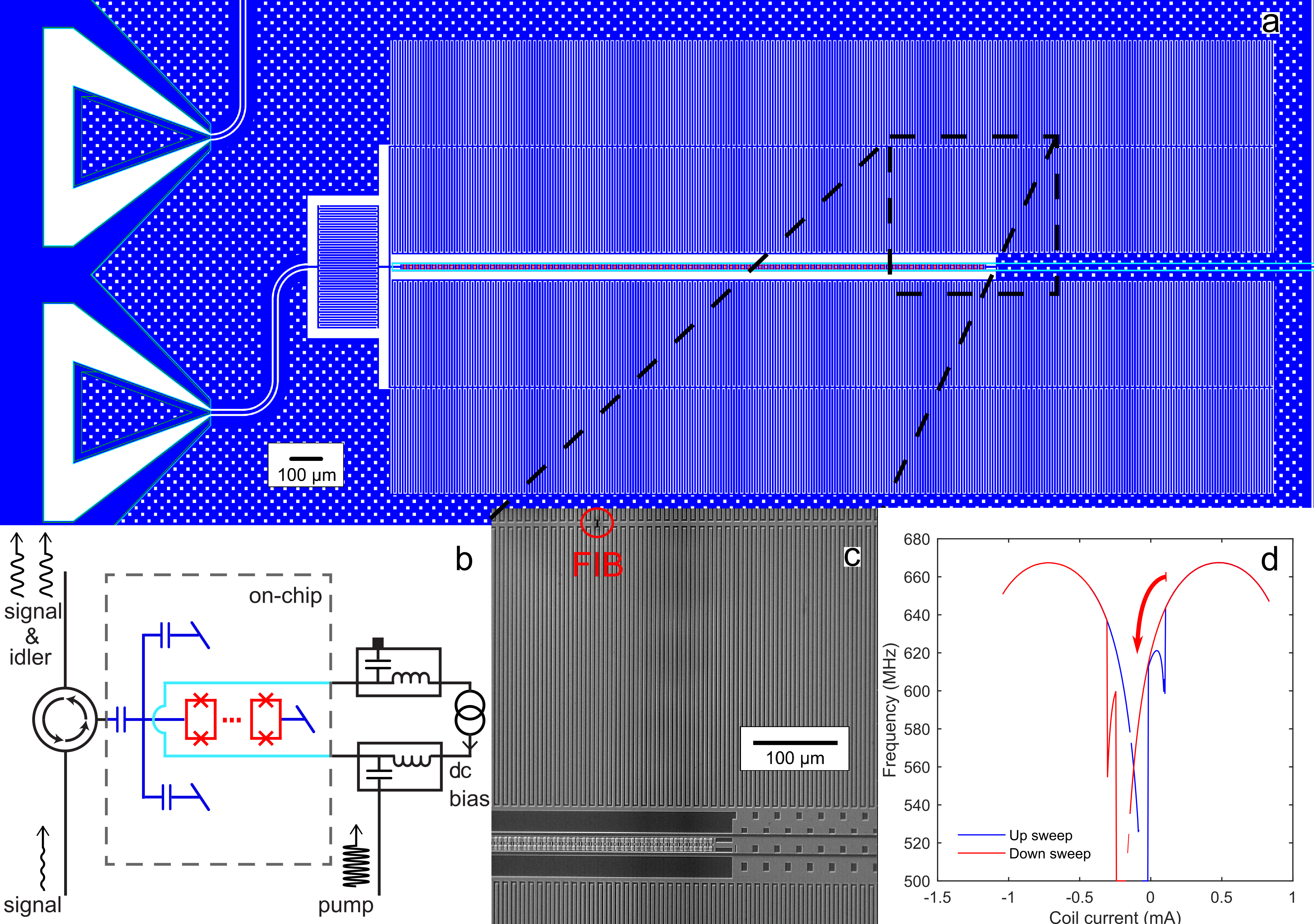}
\caption{Layout and measurement of JPA. (a) Mask drawing showing the whole JPA with the SQUID array in red and an on-chip flux bias line (FBL) in light blue. The wiring layer in dark blue features bonding pads, shunt and coupling capacitors, and the ground plane. The holes in the ground plane trap magnetic flux. (b) Simplified measurement setup with the JPA enclosed in the dashed box. The FBL carries both a pump tone and a dc current used to bias the SQUIDs and, thus, modify the resonance frequency of the device. The probe tone enters through a circulator and three-wave mixing within the JPA allows the pump tone to produce signal and idler photons according to $f_\mathrm{pump} = f_\mathrm{probe} + f_\mathrm{idler}$. A second circulator (not shown) further enhances the isolation from a subsequent cascode of post-amplifiers. (c) Scanning electron micrograph showing a part of the SQUID array and the interdigital shunt capacitor cut with a focused ion beam to fine-tune the resonance. A close-up of the other end of the array is shown in the Supplement. (d) Tuneability of the resonance frequency of Device A. Shown are a single up sweep and a down sweep of current measured with the pump off. The arrow indicates a deterministic branch selected for subsequent measurements.}
\label{fig:mask_drawing}
\end{figure*}

The devices [Fig.~\ref{fig:mask_drawing}(a)] incorporate an on-chip flux bias line (FBL) on a dedicated superconductive layer. Among many solutions for FBLs \cite{yamamoto_flux-driven_2008, sandberg_tuning_2008, krantz_investigation_2013, svensson_period-tripling_2017, mutus_design_2013, zhou_high-gain_2014}, our implementation has the advantage that the dc bias can be routed as a twisted pair through the cryostat while the rf pump tone propagates along the same on-chip conductor. A simplified wiring schematic for the devices is shown in Fig.~\ref{fig:mask_drawing}(b) and details are presented in the Supplement.

The devices are fabricated with an improved process that has been reported in detail in Ref. \onlinecite{gronberg_side-wall_2017}. In short, a high-resistivity silicon wafer is first cleaned of thermal oxides. Following that, the Josephson junctions are implemented with a niobium tri-layer (\mbox{Nb/Al-Al$_2$O$_3$/Nb}) with thicknesses 100~nm~/~10~nm~/~100~nm. The tri-layer is etched to a strip geometry and following the SWAPS process the sidewalls are passivated with PECVD silicon dioxide, as shown in a scanning electron micrograph in the Supplement. Crucially, the passivation step leaves no residual dielectric layer to the device area outside the junctions. The next step is the deposition of 120~nm of niobium for the main wiring layer. The Josephson junctions form wherever this layer crosses the tri-layer strips. Because of the need for the on-chip FBL, we add a thin, 40-nm insulating layer of ALD Al$_2$O$_3$ with wet-etched contact holes. Finally, the FBL and some superconducting cross-overs are defined from 120~nm of niobium deposited as the topmost layer. The capacitors of the previous JPA design~\cite{vesterinen_lumped-element_2017} were formed from parallel plates separated by a silicon dioxide layer, but here we use interdigitated fingers where the participation ratio of the lossy dielectrics is dramatically lower. The initial maximum resonance frequency of the devices at zero applied magnetic flux is about 500~MHz. We fine-tune it to a higher value by removing a part of the shunt capacitance with a focused ion beam~(FIB) [Fig.~\ref{fig:mask_drawing}(c)].

Two nominally identical devices A and B were prepared with the maximal resonance frequency targeted on 650~MHz, in order to make flux pumping feasible at the operating frequency of 600~MHz. The chip containing the JPA is placed on a holder and inside an aluminum-Amumetal 4K magnetic shield both of which are thermalized to the mixing stage of a dry dilution refrigerator. Prior to pumping, an initial characterization of the device takes place. It comprises the study of the small-signal response with a vector network analyzer (VNA) while tuning the resonance frequency with a magnetic flux, generated by a dc current applied to the FBL. A fit to the recorded reflection coefficient of the JPA allows us to determine the resonance frequency. Device A shows excellent reproducible tuneability between 520~MHz and 667~MHz [Fig.~\ref{fig:mask_drawing}(d)] from which can be concluded that the SQUID array has a relatively homogeneous magnetic flux bias. Only a slight hysteresis occurs below 640~MHz probably in part because of the geometric inductances \cite{pogorzalek_hysteretic_2017}. In contrast, the previous generation amplifier had very irregular frequency response and only two viable operating points \cite{vesterinen_lumped-element_2017}. There are several factors that could play a role in the observed improvement. First, a ground plane with flux-trapping holes \cite{chiaro_dielectric_2016} has been added to the device layout. Second, we have increased the size of the SQUID loops by 65~\% to $2.1\times 4.3~\mu\mathrm{m}^2$ so that less dc current is needed in the FBL. Finally, we have paid special attention to using non-magnetic materials in close proximity to the JPA chip.

\section{Gain and noise in the non-degenerate mode}

\begin{figure}[htpb]
\centering
\includegraphics[width=\hsize]{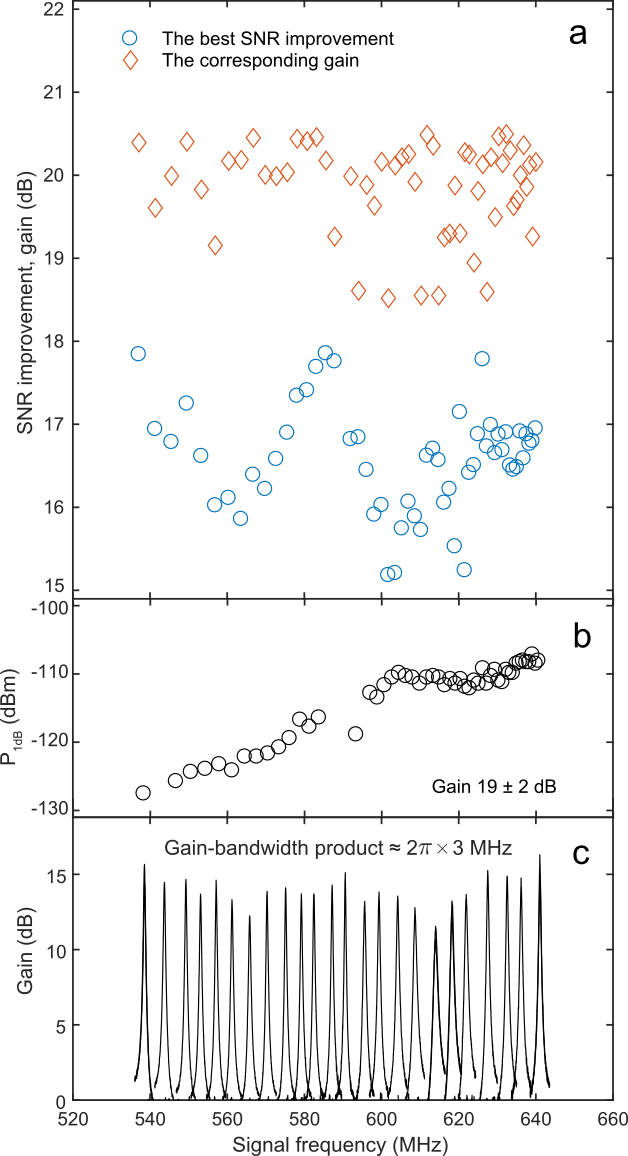}
\caption{Systematic characterization of Device A at 10~mK. At each operating point determined by a static flux bias current, the pump power is optimized first and the gain is then adjusted with the pump frequency. (a) The best signal-to-noise ratio improvement and corresponding gain while limiting to gains under 20.5~dB. The quantities are in respect to a reference that is measured with the JPA detuned and the pump switched off. (b) 1-dB compression points of gain at a probe offset of \mbox{-10}~kHz from the halved pump frequency where the gain is maximal. At around 590~MHz the pumping was unsuccessful. (c) Lorentzian gain profiles as a function of the signal frequency at 24 selected operating points. The target gain of each measurement was 15~dB.}
\label{fig:autopump}
\end{figure}

The JPA operating points defined by the triplet of the dc bias current, the associated flux pump frequency, and pump power are optimized and characterized by an automated procedure described in the Supplement. To study the gain and signal-to-noise ratio (SNR), we apply a weak probe tone at an offset of \mbox{-10}~kHz from the halved pump frequency where the JPA gain is maximal. The probe power is set to \mbox{-146}~dBm (\mbox{-136}~dBm) at frequencies below (above) 580~MHz, to adjust accordingly to the dynamic range. In the data of Fig.~\ref{fig:autopump}(a), the SNR is optimized at each static-flux operating point, while constraining the maximum gain to 20.5~dB. A gain of \mbox{18.5--20.5}~dB is attained and the SNR improves by \mbox{15--18}~dB, as compared to the unity-gain reference where the noise floor is set by the HEMT post-amplifier. The independently measured noise added by the HEMT is \mbox{10--13}~K. The saturation of the JPA is investigated at the discovered operating points by varying the probe power [Fig.~\ref{fig:autopump}(b)] and it is found that the lower limit of \mbox{-120}~dBm, required for calorimeter readout, is easily surpassed by about 10~dB at 600~MHz. After increasing the pump frequency slightly to lower the gain to 15~dB, a gain-bandwidth product of $2\pi$~$\times$~$3$~MHz is measured [Fig.~\ref{fig:autopump}(c)]. It is sufficient for the detection of the thermal transients of the nano-calorimeter.

We investigate the potential range of the operating temperatures of Device A by setting a gain of 20~dB and measuring the JPA response with the VNA while heating the system. The frequency and power of the flux pump are kept at the fixed values. We attribute changes in the gain to the temperature dependence of the JPA resonance frequency. Raising the temperature from 40~mK to 80~mK (400~mK) increases the gain by 1~dB (5~dB). The gain stays above 20~dB below 1.0~K, and even at 1.5~K there is still a significant 16~dB of gain. The equivalent shift of the JPA resonance is $\pm$~200~kHz, or $\pm$~0.04~\%. We note that these observations result from an interplay of several temperature-dependent quantities such as the permittivity of the Al$_2$O$_3$ coating of the JPA, the junction critical current, and the kinetic inductance of niobium \cite{annunziata_tunable_2010}. 
\begin{figure}[htpb]
\centering
\includegraphics[width=\hsize]{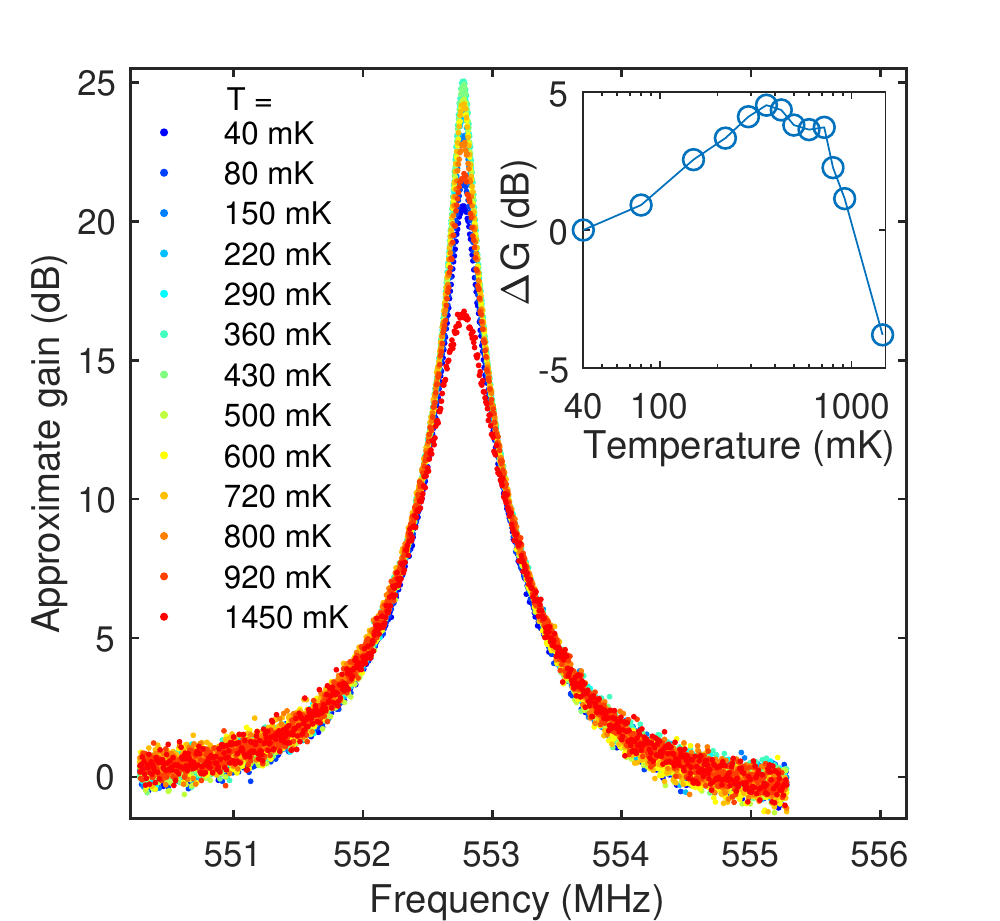}
\caption{Approximate gain profiles for Device A at 553~MHz at varying temperatures of the sample stage starting from a gain of 20~dB at 40~mK. In the absence of reference data, the recorded signal is normalized to the edges of the peak. The pump power or frequency are not re-optimized during heating proving the temperature robustness of the amplifier. Inset: change in gain from 40~mK, as a function of temperature.}
\label{fig:gain_vs_freq_vs_temp}
\end{figure}

\begin{figure}[htpb]
\centering
\includegraphics[width=\hsize]{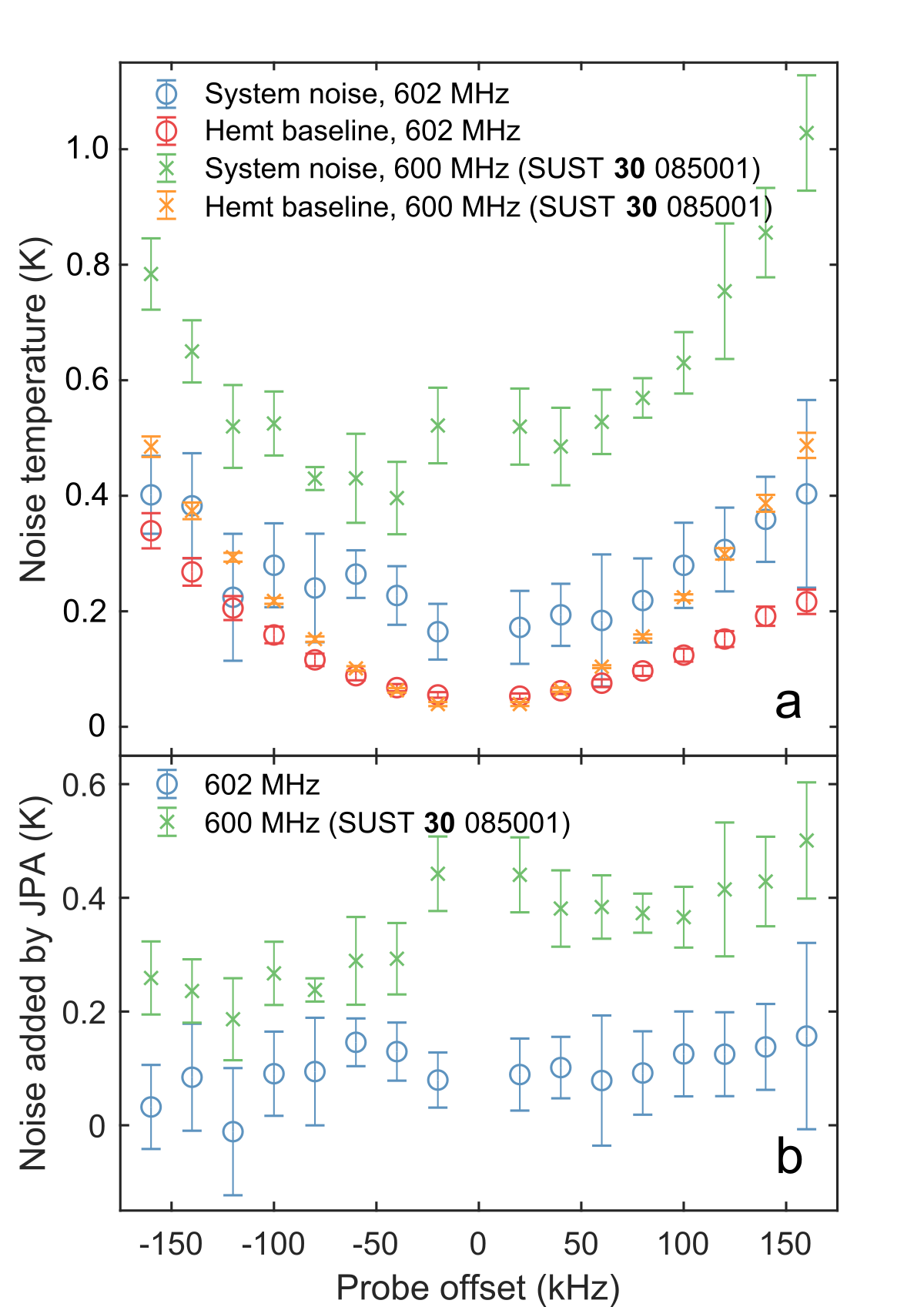}
\caption{Noise temperature of Device B at 600 MHz determined with the Y-factor method using a calibrated noise source. Data from Ref.~\onlinecite{vesterinen_lumped-element_2017} are added for comparison. (a) System noise temperature $T_\mathrm{sys}$ as a function of the probe offset from the halved pump frequency. The separated contribution of the HEMT amplifier to $T_\mathrm{sys}$ is the smallest at zero offset where the JPA gain is maximal. (b) The noise added by the JPA is the difference between the system noise temperature and the part added by HEMT as well as the input noise: 30~mK in recent experiments and 40~mK in Ref.~\onlinecite{vesterinen_lumped-element_2017}.}
\label{fig:noise_temp}
\end{figure}

To better estimate the noise added by the JPA, the system noise temperature is determined with the Y-factor method \cite{kinion_superconducting_2011}. Essentially, the power spectral density at the output of the JPA is surveyed with a spectrum analyzer while the JPA is subject to an impedance-matched resistive noise source with a controlled temperature. The source temperature is varied between 59~mK and 852~mK independently of the JPA temperature that is held constant at 30~mK with closed-loop control. The measurements are carried out with the Device B and the full setup is shown in the Supplement. The system noise temperature, referenced to the JPA input, takes its minimum value of 165~mK close to the halved pump frequency [Fig.~\ref{fig:noise_temp}(a)]. Since the noise added by the JPA is reasonably independent of the offset [Fig.~\ref{fig:noise_temp}(b)], we may take the average with the inverted variances as weights. This yields a noise estimate of 105~mK~$\pm$~9~mK. Making the comparison to Ref.~\onlinecite{vesterinen_lumped-element_2017}, we note that the added noise has approximately halved. Also notable is that the noise is no longer at an elevated level at frequencies close to the gain maximum, which may be because the rf pump tone has been moved away from it to the double frequency.

\section{Degenerate gain}

In a dispersive sensor, such as the nano-calorimeters of Ref.~\onlinecite{viisanen_incomplete_2015}, it is possible to choose the excitation and readout in such a way that the rf carrier and the information-carrying signal are in different quadratures. In this situation, it is possible to utilize squeezing to de-amplify the carrier and to amplify the signal. This may be used to effectively increase the dynamic range of the later stages of the readout. Ideally, the amplification does not add any noise if a JPA is used to perform the squeezing.

\begin{figure}[htpb]
\centering
\includegraphics[width=\hsize]{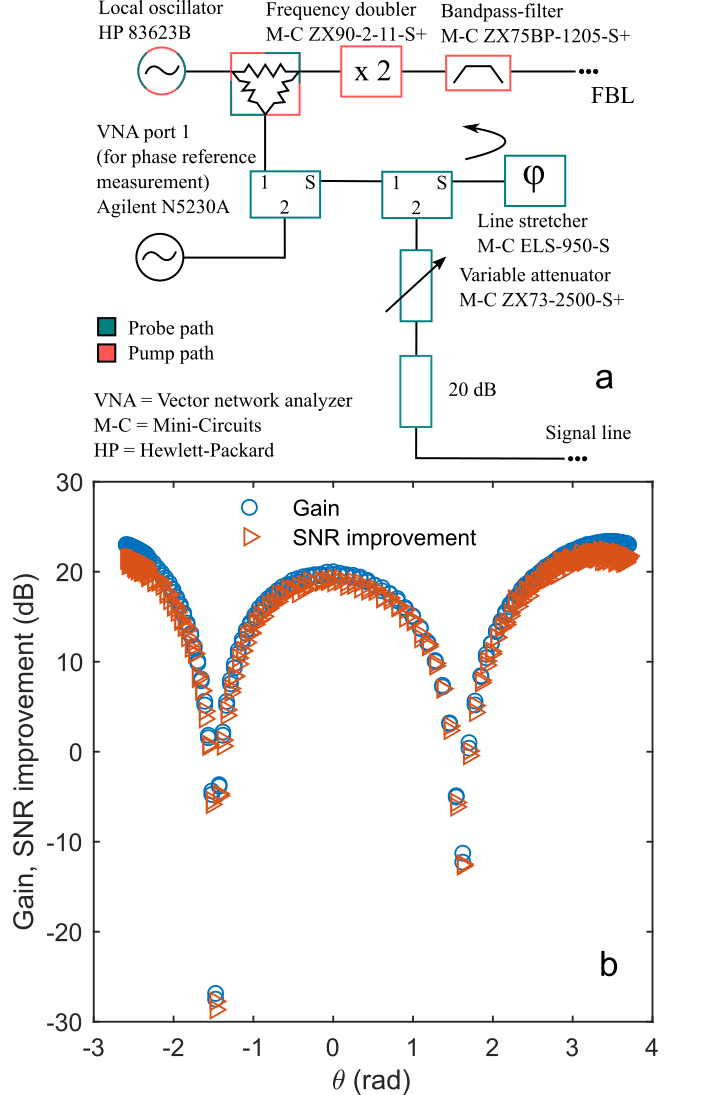}
\caption{Device B in the degenerate mode at an operating temperature of 30~mK. The probe frequency reads $f_\mathrm{probe}$ = $\frac{1}{2}f_\mathrm{pump}$ = $f_\mathrm{idler}$ = 601.85~MHz. (a) A frequency doubler generates the phase-locked pump tone. The probe is directed to a voltage-controlled electronic line stretcher (ELS) to adjust the phase difference between the probe and the pump. Afterwards, the VNA is used to calibrate the phase versus the control voltage. (b) Gain (circles) and SNR improvement (triangles) as a function of the calibrated phase. The latter is determined by comparing to a measurement where the JPA is detuned with applied magnetic flux. The observed difference of 3~dB between the gain maxima is likely caused by varying pump power, due to voltage dependence in the reflection coefficient of the ELS: a signal leaking backwards to the input of the doubler interferes with the signal following the direct path.}
\label{fig:phase_sensitive}
\end{figure}

To observe squeezing in the degenerate mode of the JPA, where the pump is exactly at twice the probe frequency, a brief experiment is carried out with the Device B. Using a single rf source, the pump is synthesized with a frequency doubler and the relative phase $\theta$ of the probe is controlled as shown in Fig.~\ref{fig:phase_sensitive}(a). The JPA gain as a function of $\theta$ is $\pi$-periodic as expected, and amplification and de-amplification alternate. The JPA is thus capable of squeezing~\cite{zhong_squeezing_2013}. However, here we do not attempt a proper investigation of the quadratures to determine how much vacuum squeezing is attainable.

\clearpage

\section{Conclusion}
We have overcome all the major issues discovered with the previous-generation JPA \cite{vesterinen_lumped-element_2017}, and we have attained robust performance over a wide frequency range without sacrificing bandwidth, gain, or dynamic range. Additionally, we have minimized the amount of TLS-hosting dielectrics and their participation ratio with the SWAPS \cite{gronberg_side-wall_2017} fabrication process. This renders the JPA relatively insensitive to temperature, facilitating its use at variable mK-temperatures as only minor corrections to the JPA operating parameters are needed. Furthermore, the JPA is now flux-pumped at twice the readout frequency of the nano-calorimeter, easing the filtering required to avoid back-action such as residual heating by the pump. As an important step towards sub-GHz sensor readout at a fidelity beyond the standard quantum limit \cite{caves_quantum_1982}, squeezing has been observed in the degenerate mode of the JPA.

The noise added by the JPA has decreased to 105~mK at 600~MHz. This can largely be attributed to better isolation from the HEMT post-amplifier by means of an additional circulator. However, the added noise remains at an elevated level with respect to the lower bound set by the input thermal noise at 30~mK, likely due to poorly thermalized attenuators \cite{yeh_microwave_2017} or a fundamental limit of the amplifier itself. The noise performance also remains inferior to that of a microstrip SQUID amplifier with a reported noise temperature of 48~mK at 612~MHz \cite{kinion_superconducting_2011}. Yet, we have demonstrated a sub-GHz JPA that is well suited for integration into a nano-calorimetry \cite{viisanen_incomplete_2015} or -bolometry \cite{govenius_detection_2016} apparatus and we will pursue the latter goal in a future experiment.

\section*{Acknowledgments}
We thank Paula Holmlund for sample preparation, Harri Pohjonen for help with lithographic masks, and Mikko Kiviranta for operating the FIB. We acknowledge the fruitful discussions with Olli-Pentti Saira, Roope Kokkoniemi, Mikko M\"ott\"onen, and Jukka Pekola. This work was performed as part of the Academy of Finland Centre of Excellence program (projects 284594, 312059, 251748, and 284621). The work also received funding from Academy of Finland project QuMOS (project numbers 288907 and 287768), Future Makers Funding Program by Technology Industries of Finland Centennial Foundation and Jane and Aatos Erkko Foundation.

\noappendix
\clearpage
\pagestyle{plain}
\newcommand{\newblock}{}
\bibliographystyle{iopart-num}
\bibliography{jpa_article_2018}

\providecommand{\newblock}{}
\begin{thebibliography}{10}
\expandafter\ifx\csname url\endcsname\relax
  \def\url#1{{\tt #1}}\fi
\expandafter\ifx\csname urlprefix\endcsname\relax\def\urlprefix{URL }\fi
\providecommand{\eprint}[2][]{\url{#2}}

\bibitem{bradley_microwave_2003}
Bradley R, Clarke J, Kinion D, Rosenberg L~J, van Bibber K, Matsuki S, Mück M
  and Sikivie P 2003 {\em Reviews of Modern Physics\/} {\bf 75} 777

\bibitem{asztalos_squid-based_2010}
Asztalos S~J, Carosi G, Hagmann C, Kinion D, van Bibber K, Hotz M, Rosenberg
  L~J, Rybka G, Hoskins J, Hwang J, Sikivie P, Tanner D~B, Bradley R and Clarke
  J 2010 {\em Physical Review Letters\/} {\bf 104} 041301

\bibitem{kenany_design_2017}
Kenany S~A, Anil M~A, Backes K~M, Brubaker B~M, Cahn S~B, Carosi G, Gurevich
  Y~V, Kindel W~F, Lamoreaux S~K, Lehnert K~W, Lewis S~M, Malnou M, Palken D~A,
  Rapidis N~M, Root J~R, Simanovskaia M, Shokair T~M, Urdinaran I, van Bibber
  K~A and Zhong L 2017 {\em Nuclear Instruments and Methods in Physics Research
  Section A: Accelerators, Spectrometers, Detectors and Associated Equipment\/}
  {\bf 854} 11--24

\bibitem{bergeal_phase-preserving_2010}
Bergeal N, Schackert F, Metcalfe M, Vijay R, Manucharyan V~E, Frunzio L, Prober
  D~E, Schoelkopf R~J, Girvin S~M and Devoret M~H 2010 {\em Nature\/} {\bf 465}
  64--68

\bibitem{abdo_josephson_2014}
Abdo B, Sliwa K, Shankar S, Hatridge M, Frunzio L, Schoelkopf R and Devoret M
  2014 {\em Physical Review Letters\/} {\bf 112} 167701

\bibitem{obrien_towards_2016}
O'Brien K, Macklin C, Hover D, Schwartz M~E, Bolkhovsky V, Zhang X, Oliver W~D
  and Siddiqi I 2016 Towards quantum-noise limited multiplexed microwave
  readout of qubits {\em 2016 {IEEE} {MTT}-{S} {International} {Microwave}
  {Symposium} ({IMS})\/} pp 1--3

\bibitem{devoret_superconducting_2013}
Devoret M~H and Schoelkopf R~J 2013 {\em Science\/} {\bf 339} 1169--1174

\bibitem{calusine_analysis_2018}
Calusine G, Melville A, Woods W, Das R, Stull C, Bolkhovsky V, Braje D, Hover
  D, Kim D~K, Miloshi X, Rosenberg D, Sevi A, Yoder J~L, Dauler E and Oliver
  W~D 2018 {\em Applied Physics Letters\/} {\bf 112} 062601

\bibitem{clark_observation_2016}
Clark J~B, Lecocq F, Simmonds R~W, Aumentado J and Teufel J~D 2016 {\em Nature
  Physics\/} {\bf 12} 683--687

\bibitem{yurke_observation_1988}
Yurke B, Kaminsky P~G, Miller R~E, Whittaker E~A, Smith A~D, Silver A~H and
  Simon R~W 1988 {\em Physical Review Letters\/} {\bf 60} 764--767

\bibitem{castellanos-beltran_widely_2007}
Castellanos-Beltran M~A and Lehnert K~W 2007 {\em Applied Physics Letters\/}
  {\bf 91} 083509

\bibitem{yamamoto_flux-driven_2008}
Yamamoto T, Inomata K, Watanabe M, Matsuba K, Miyazaki T, Oliver W~D, Nakamura
  Y and Tsai J~S 2008 {\em Applied Physics Letters\/} {\bf 93} 042510

\bibitem{mutus_design_2013}
Mutus J~Y, White T~C, Jeffrey E, Sank D, Barends R, Bochmann J, Chen Y, Chen Z,
  Chiaro B, Dunsworth A, Kelly J, Megrant A, Neill C, O'Malley P~J~J, Roushan
  P, Vainsencher A, Wenner J, Siddiqi I, Vijay R, Cleland A~N and Martinis J~M
  2013 {\em Applied Physics Letters\/} {\bf 103} 122602

\bibitem{eichler_controlling_2014}
Eichler C and Wallraff A 2014 {\em EPJ Quantum Technology\/} {\bf 1} 2

\bibitem{mutus_strong_2014}
Mutus J~Y, White T~C, Barends R, Chen Y, Chen Z, Chiaro B, Dunsworth A, Jeffrey
  E, Kelly J, Megrant A, Neill C, O'Malley P~J~J, Roushan P, Sank D,
  Vainsencher A, Wenner J, Sundqvist K~M, Cleland A~N and Martinis J~M 2014
  {\em Applied Physics Letters\/} {\bf 104} 263513

\bibitem{eichler_quantum-limited_2014}
Eichler C, Salathe Y, Mlynek J, Schmidt S and Wallraff A 2014 {\em Physical
  Review Letters\/} {\bf 113} 110502

\bibitem{vissers_low-noise_2016}
Vissers M~R, Erickson R~P, Ku H~S, Vale L, Wu X, Hilton G~C and Pappas D~P 2016
  {\em Applied Physics Letters\/} {\bf 108} 012601

\bibitem{mallet_quantum_2011}
Mallet F, Castellanos-Beltran M~A, Ku H~S, Glancy S, Knill E, Irwin K~D, Hilton
  G~C, Vale L~R and Lehnert K~W 2011 {\em Physical Review Letters\/} {\bf 106}
  220502

\bibitem{fedorov_displacement_2016}
Fedorov K~G, Zhong L, Pogorzalek S, Eder P, Fischer M, Goetz J, Xie E,
  Wulschner F, Inomata K, Yamamoto T, Nakamura Y, Di~Candia R, Las~Heras U,
  Sanz M, Solano E, Menzel E~P, Deppe F, Marx A and Gross R 2016 {\em Physical
  Review Letters\/} {\bf 117} 020502

\bibitem{bienfait_magnetic_2017}
Bienfait A, Campagne-Ibarcq P, Kiilerich A, Zhou X, Probst S, Pla J, Schenkel
  T, Vion D, Esteve D, Morton J, Moelmer K and Bertet P 2017 {\em Physical
  Review X\/} {\bf 7} 041011

\bibitem{clerk_introduction_2010}
Clerk A~A, Devoret M~H, Girvin S~M, Marquardt F and Schoelkopf R~J 2010 {\em
  Reviews of Modern Physics\/} {\bf 82} 1155--1208

\bibitem{vesterinen_lumped-element_2017}
Vesterinen V, Saira O~P, Räisänen I, Möttönen M, Grönberg L, Pekola J and
  {Juha Hassel} 2017 {\em Superconductor Science and Technology\/} {\bf 30}
  085001

\bibitem{gasparinetti_fast_2015}
Gasparinetti S, Viisanen K~L, Saira O~P, Faivre T, Arzeo M, Meschke M and
  Pekola J~P 2015 {\em Physical Review Applied\/} {\bf 3} 014007

\bibitem{viisanen_incomplete_2015}
Viisanen K~L, Suomela S, Gasparinetti S, Saira O~P, Ankerhold J and Pekola J~P
  2015 {\em New Journal of Physics\/} {\bf 17} 055014

\bibitem{govenius_microwave_2014}
Govenius J, Lake R~E, Tan K~Y, Pietilä V, Julin J~K, Maasilta I~J, Virtanen P
  and Möttönen M 2014 {\em Physical Review B\/} {\bf 90} 064505

\bibitem{govenius_detection_2016}
Govenius J, Lake R~E, Tan K~Y and Möttönen M 2016 {\em Physical Review
  Letters\/} {\bf 117} 030802

\bibitem{karasik_demonstration_2011}
Karasik B~S and Cantor R 2011 {\em Applied Physics Letters\/} {\bf 98} 193503

\bibitem{suzuki_performance_2014}
Suzuki T, Khosropanah P, Hijmering R~A, Ridder M, Schoemans M, Hoevers H and
  Gao J~R 2014 {\em IEEE Transactions on Terahertz Science and Technology\/}
  {\bf 4} 171--178

\bibitem{penfold-fitch_microwave_2017}
Penfold-Fitch Z~V, Sfigakis F and Buitelaar M~R 2017 {\em Physical Review
  Applied\/} {\bf 7} 054017

\bibitem{oconnell_microwave_2008}
O’Connell A~D, Ansmann M, Bialczak R~C, Hofheinz M, Katz N, Lucero E,
  McKenney C, Neeley M, Wang H, Weig E~M, Cleland A~N and Martinis J~M 2008
  {\em Applied Physics Letters\/} {\bf 92} 112903

\bibitem{pappas_two_2011}
Pappas D~P, Vissers M~R, Wisbey D~S, Kline J~S and Gao J 2011 {\em IEEE
  Transactions on Applied Superconductivity\/} {\bf 21} 871--874

\bibitem{gronberg_side-wall_2017}
Grönberg L, Kiviranta M, Vesterinen V, Lehtinen J, Simbierowicz S, Luomahaara
  J, Prunnila M and Hassel J 2017 {\em Superconductor Science and Technology\/}
  {\bf 30} 125016

\bibitem{kiviranta_multilayer_2016}
Kiviranta M, Brandel O, Grönberg L, Kunert J, Linzen S, Beev N, May T and
  Prunnila M 2016 {\em IEEE Transactions on Applied Superconductivity\/} {\bf
  26} 1--5

\bibitem{lahteenmaki_dynamical_2013}
Lähteenmäki P, Paraoanu G~S, Hassel J and Hakonen P~J 2013 {\em Proceedings
  of the National Academy of Sciences\/} {\bf 110} 4234--4238

\bibitem{devoret_introduction_2016}
Devoret M and Roy A 2016 {\em Comptes Rendus Physique\/} {\bf 17} 740--755

\bibitem{noauthor_supplementary_nodate}
Supplementary information \urlprefix\url{Supplementary information available
  online at.}

\bibitem{sandberg_tuning_2008}
Sandberg M, Wilson C~M, Persson F, Bauch T, Johansson G, Shumeiko V, Duty T and
  Delsing P 2008 {\em Applied Physics Letters\/} {\bf 92} 203501

\bibitem{krantz_investigation_2013}
Krantz P, Reshitnyk Y, Wustmann W, Bylander J, Gustavsson S, Oliver W~D, Duty
  T, Shumeiko V and Delsing P 2013 {\em New Journal of Physics\/} {\bf 15}
  105002

\bibitem{svensson_period-tripling_2017}
Svensson I~M, Bengtsson A, Krantz P, Bylander J, Shumeiko V and Delsing P 2017
  {\em Physical Review B\/} {\bf 96} 174503

\bibitem{zhou_high-gain_2014}
Zhou X, Schmitt V, Bertet P, Vion D, Wustmann W, Shumeiko V and Esteve D 2014
  {\em Physical Review B\/} {\bf 89} 214517

\bibitem{pogorzalek_hysteretic_2017}
Pogorzalek S, Fedorov K~G, Zhong L, Goetz J, Wulschner F, Fischer M, Eder P,
  Xie E, Inomata K, Yamamoto T, Nakamura Y, Marx A, Deppe F and Gross R 2017
  {\em Physical Review Applied\/} {\bf 8} 024012

\bibitem{chiaro_dielectric_2016}
Chiaro B, Megrant A, Dunsworth A, Chen Z, Barends R, Campbell B, Chen Y, Fowler
  A, Hoi I~C, Jeffrey E, Kelly J, Mutus J, Neill C, O'Malley P~J~J, Quintana C,
  Roushan P, Sank D, Vainsencher A, Wenner J, White T~C and Martinis J~M 2016
  {\em Superconductor Science and Technology\/} {\bf 29} 104006

\bibitem{annunziata_tunable_2010}
Annunziata A~J, Santavicca D~F, Frunzio L, Catelani G, Rooks M~J, Frydman A and
  Prober D~E 2010 {\em Nanotechnology\/} {\bf 21} 445202

\bibitem{kinion_superconducting_2011}
Kinion D and Clarke J 2011 {\em Applied Physics Letters\/} {\bf 98} 202503

\bibitem{zhong_squeezing_2013}
Zhong L, Menzel E~P, Di~Candia R, Eder P, Ihmig M, Baust A, Haeberlein M,
  Hoffmann E, Inomata K, Yamamoto T, Nakamura Y, Solano E, Deppe F, Marx A and
  Gross R 2013 {\em New Journal of Physics\/} {\bf 15} 125013

\bibitem{caves_quantum_1982}
Caves C~M 1982 {\em Physical Review D\/} {\bf 26} 1817--1839

\bibitem{yeh_microwave_2017}
Yeh J~H, LeFebvre J, Premaratne S, Wellstood F~C and Palmer B~S 2017 {\em
  Journal of Applied Physics\/} {\bf 121} 224501

\end{thebibliography}


\providecommand{\newblock}{}
\begin{thebibliography}{1}
\expandafter\ifx\csname url\endcsname\relax
  \def\url#1{{\tt #1}}\fi
\expandafter\ifx\csname urlprefix\endcsname\relax\def\urlprefix{URL }\fi
\providecommand{\eprint}[2][]{\url{#2}}

\bibitem{vesterinen_lumped-element_2017}
Vesterinen V, Saira O~P, Räisänen I, Möttönen M, Grönberg L, Pekola J and
  {Juha Hassel} 2017 {\em Superconductor Science and Technology\/} {\bf 30}
  085001

\bibitem{krantz_investigation_2013}
Krantz P, Reshitnyk Y, Wustmann W, Bylander J, Gustavsson S, Oliver W~D, Duty
  T, Shumeiko V and Delsing P 2013 {\em New Journal of Physics\/} {\bf 15}
  105002

\bibitem{manucharyan_microwave_2007}
Manucharyan V~E, Boaknin E, Metcalfe M, Vijay R, Siddiqi I and Devoret M 2007
  {\em Physical Review B\/} {\bf 76} 014524

\end{thebibliography}

\end{document}


\title[]{Supplement for article "Flux-driven Josephson parametric amplifier for sub-GHz frequencies fabricated with side-wall passivated spacer junction technology"}

\author{Slawomir Simbierowicz$^1$, Visa Vesterinen$^{1,2}$, Leif Gr\"onberg$^1$, Janne Lehtinen$^1$, Mika Prunnila$^1$ and Juha Hassel$^1$}
\address{$^1$ VTT Technical Research Centre of Finland Ltd \& QTF Centre of Excellence, P.O.Box 1000, FI-02044 VTT, FINLAND}
\address{$^2$ QCD Labs, COMP Centre of Excellence, Department of Applied Physics, Aalto University, P.O. Box 13500, 00076 Aalto, Finland}
\ead{slawomir.simbierowicz@vtt.fi}
\vspace{10pt}
\begin{indented}
\item[]May 2018
\end{indented}

\maketitle

\section*{Dynamic range}
The limitation of dynamical range at low frequency is circumvented by adding 200 SQUIDs in series, enabled by the expression for the bifurcation power of a parametric current pump \cite{vesterinen_lumped-element_2017}:
\begin{equation}
P_\mathrm{bf} = \frac{N^2 \chi^2 Q \phi_0^2 \omega_0^2}{32 \sqrt{3} \eta Q_\mathrm{e} Z_\mathrm{LC} } \mathrm{ ,}
\end{equation}
where $N$ is the number of Josephson elements in series, $\omega_0/(2\pi)$ is the drive frequency of the device, $Q$ and $Q_\mathrm{e}$ are the total and external quality factors, and $\chi = \frac{16}{\sqrt{3}}Q^{-1}\eta^{-1}$. The quantity $\eta = 1/(1 + L_\mathrm{geom}^{(1)} / L_\mathrm{J}^{(1)})$ is controlled with the geometric and Josephson inductances. Finally, $\phi_0 = \hbar/(2e)$ is the reduced flux quantum. Our devices were designed using a ratio of $L_\mathrm{geom}^{(1)} / L_\mathrm{J}^{(1)} = 0.30$ and a characteristic impedance of $Z_\mathrm{LC} = \sqrt{L/C} = 8.8~\Omega$, where $L = L_\mathrm{geom}^{(1)} + L_\mathrm{J}^{(1)}$ and $C$ is the total capacitance of the resonator. Determining quality factors from experiments performed on Device A, we get $Q_\mathrm{e} = 230$ and $Q_\mathrm{i} = 1771$ at 601~MHz. Then, $\chi = 0.06$ and $P_\mathrm{bf}$ = \mbox{-92.9}~dBm which is 1~dB higher than in Ref.~\onlinecite{vesterinen_lumped-element_2017} and sufficient for amplifying nano-calorimeter signals.

\section*{Junction structure}

\begin{figure*}[htpb]
\centering
\includegraphics[width=\hsize]{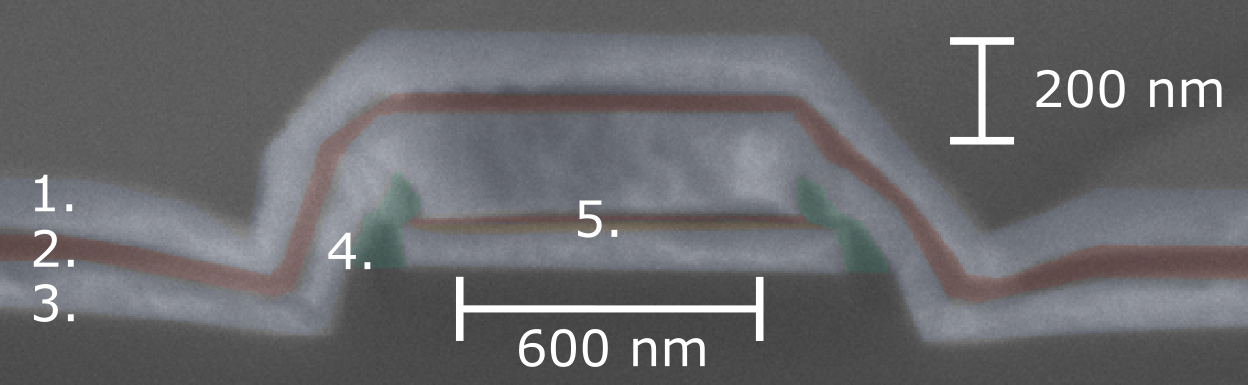}
\caption{False colored scanning electron micrograph showing a cross-section of a test junction sharing the layer composition of the JPAs. A high-resistivity silicon substrate carries the structure with the following elements: 1. the niobium cross-overs and the flux bias line; 2. the insulating ALD aluminium oxide layer; 3. the niobium ground plane and wiring layer; 4. a silicon dioxide spacer used for the passivation of junction side-walls; 5. the \mbox{Nb-Al/AlO$_\mathrm{x}$-Nb} trilayer containing the tunnel junctions.}
\end{figure*}\label{fig:sem_junction}

\clearpage

\section*{SQUID close-up}

\begin{figure*}[htpb]
\centering
\includegraphics[width=\hsize]{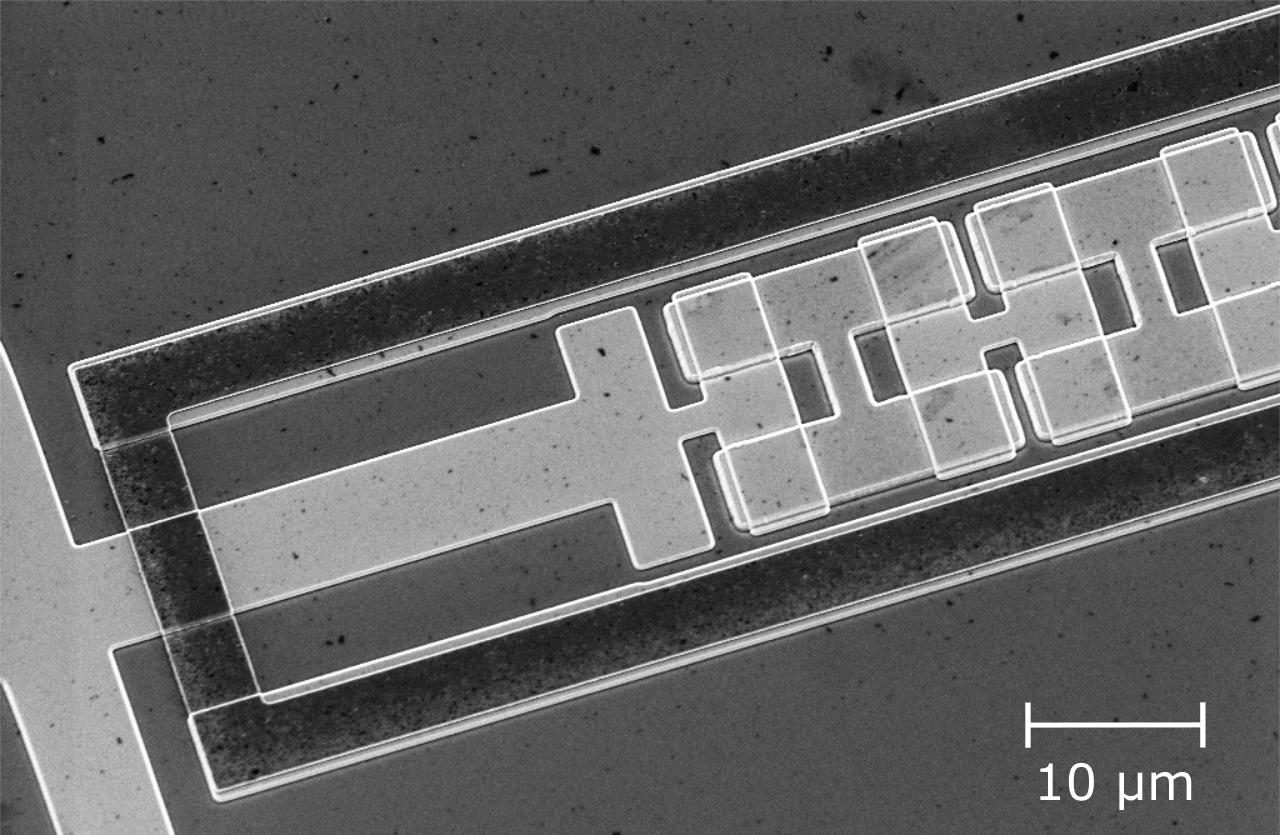}
\caption{Scanning electron micrograph showing one end of the niobium SQUID array. The flux bias line (black) can be seen crossing over the center conductor.}
\end{figure*}\label{fig:sem_squid}

\clearpage

\section*{Full cryogenic setup}
\begin{figure*}[htpb]
\centering
\includegraphics[width=\hsize]{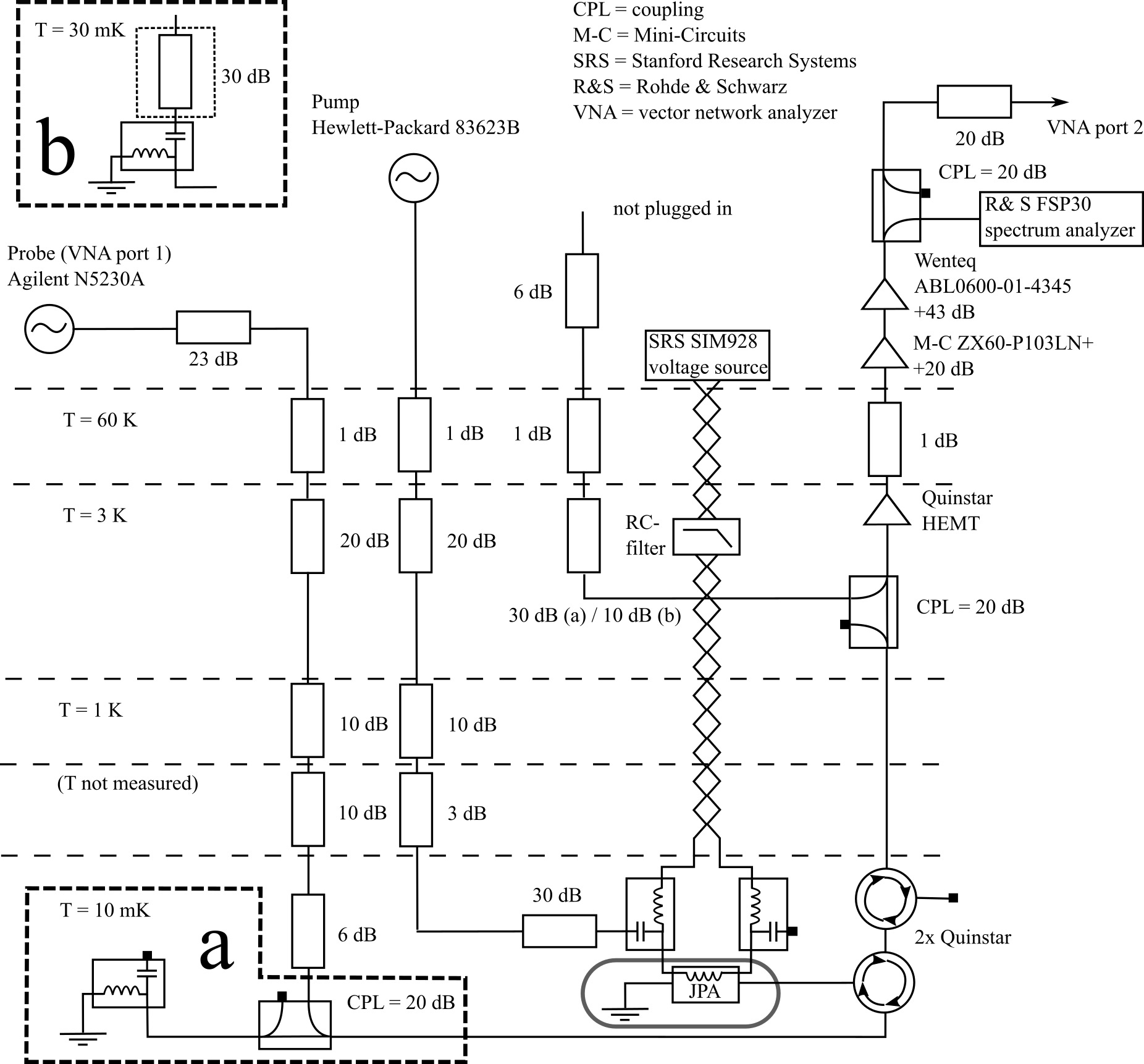}
\caption{Detailed wiring schematic. The JPA lies inside an aluminum-Amumetal 4K magnetic shield (gray line) thermalized to the mixing plate (MP) of a dilution refrigerator (BlueFors Cryogenics LD-250). A twisted pair of wires carries DC current to the on-chip coil biasing the SQUID chain on the device. The DC bias is combined with the RF pump using two bias-tees: the second one terminates the pump line. Both probe and pump RF tones are heavily attenuated in steps. Two back-to-back circulators protect the JPA from HEMT ($T_\mathrm{N}$ = \mbox{10--13}~K) back-action. Gain and noise measurements are performed using the VNA and the spectrum analyzer. The redundant RF line was previously used for cancellation of the pump tone during current pumping. All RF instruments are synchronized with a rubidium frequency standard (SRS FS725, not shown). (a) Measurement of Device A happens at 10~mK. The combination of the directional coupler and the bias-tee thermalizes the JPA to the MP. (b) Measurements for Device B are performed at 30 mK and the signal is rerouted through a noise source: a heated 30~dB attenuator (XMA Corp.) inside a copper enclosure (small dashed box) weakly thermalized to the MP.}
\end{figure*}\label{fig:full_setup}

\section*{Automated pump procedure}

Before executing the automated pumping procedure, some parameters need to be selected: the range of dc coil currents, the probe power, the probe offset from the pump, and the power window during the probe power sweep that measures the 1-dB saturation point of the amplifier. Except in the saturation measurement, the probe power should be low enough that a small change in power does not affect the gain and high enough for an adequate signal-to-noise ratio, especially when the pump has been turned off. We use a spectrum analyzer to determine SNR and gain accurately. Additionally, in order to benefit from the higher dynamic range at high signal frequencies, the probe power is boosted by 10~dB there. Choosing the parameters may require manual operation of the amplifier.

The automated script is executed at each dc coil current. It starts by fitting to S-parameter data procured by the VNA and sets the pump frequency to \mbox{-3}~MHz from twice the obtained resonance frequency. For an expected gain-bandwidth product of 3~MHz, this puts the halved pump at a frequency offset reasonably detuned from the JPA resonance. Next, the pump power is increased until the JPA goes into the mode of parametric oscillation \cite{krantz_investigation_2013} which manifests itself as a strong peak measured with the spectrum analyzer at the halved pump frequency. The power is then lowered in steps of 0.1~dBm until the oscillation disappears. We call this point the bifurcation power because of the hysteresis of the parametric oscillation. We further decrease the pump power by 0.15~dBm, arriving right below the so-called line of maximum gain \cite{manucharyan_microwave_2007}. After this, the pump frequency is increased in steps of 100~kHz to find the range of available gain while the probe trails the halved pump frequency by the predefined offset of \mbox{-10}~kHz. The spectrum analyzer is used to measure the probe power and the noise in a 20~kHz window centered at the probe frequency.

Whenever reaching a gain threshold of 20~dB, the user-defined sweep of the probe power is performed with the VNA. Similarly, at a gain threshold of 15~dB the VNA measures the gain-bandwidth product using a 5~MHz window centered at the gain maximum. To calibrate the gain, the JPA is momentarily detuned and the pump is turned off to measure the reference without changing the settings of the VNA. Finally, the script re-visits all the frequency windows on the spectrum analyzer but with the JPA detuned and the pump turned off. This establishes the references for the probe power and the noise.

\noappendix
\clearpage
\pagestyle{plain}

\bibliographystyle{iopart-num}
\bibliography{jpa_article_2018}